\documentclass[10pt,conference,letterpaper]{IEEEtran}
\IEEEoverridecommandlockouts

\usepackage{cite}
\usepackage{amsmath,amssymb,amsfonts}
\usepackage{algorithmic}
\usepackage{graphicx}
\usepackage{textcomp}
\usepackage{xcolor}
\def\BibTeX{{\rm B\kern-.05em{\sc i\kern-.025em b}\kern-.08em
    T\kern-.1667em\lower.7ex\hbox{E}\kern-.125emX}}
\usepackage{comment}

\usepackage{amsmath,amssymb,amsthm,mathrsfs}

\usepackage{mathtools}

\usepackage{colortbl}
\usepackage{booktabs}
\usepackage{tabularray}
\usepackage{multirow}

\usepackage{graphicx}
\graphicspath{{figs/}}
\DeclareGraphicsExtensions{.pdf}
\usepackage{epsfig}
\usepackage{epstopdf} 
\usepackage{rotating}
\usepackage{wrapfig}
\usepackage{enumitem}
\usepackage{lipsum}

\usepackage{tikz}
\usetikzlibrary{tikzmark}
\tikzset{mycircled/.style={circle,draw,inner sep=0.15em,line width=0.1em}}

\usepackage{algorithmic}

\usepackage{textcomp}
\usepackage{xspace}
\usepackage{float}
\usepackage{url}
\usepackage[normalem]{ulem}

\usepackage{listings}
\usepackage{flushend}
\usepackage{pifont}
\usepackage{soul}
\usepackage{balance}
\usepackage{array}
\usepackage{capt-of}
\usepackage{gensymb}
\usepackage{footnote}
\usepackage[bottom]{footmisc}
\usepackage{siunitx}

\usepackage[normalem]{ulem}
\useunder{\uline}{\ul}{}
\usepackage{diagbox}
\usepackage{arydshln}
 \usepackage[none]{hyphenat}
\usepackage{makecell}
\usepackage{hyperref}
\hypersetup{draft}
\usepackage{xurl}

\usepackage[printonlyused]{acronym}
\acrodef{ran}[RAN]{Radio Access Network}
\acrodef{ho}[HO]{Handover}
\acrodefplural{ho}[HOs]{Handovers}
\acrodef{prb}[PRB]{Physical Resource Block}
\acrodef{pdn}[PDN]{Packet Data Network}
\acrodef{pdu}[PDU]{Packet Data Unit}
\acrodef{e2e}[E2E]{end-to-end}
\acrodef{ue}[UE]{User Equipment}
\acrodef{qoe}[QoE]{Quality of Experience}
\acrodef{mme}[MME]{Mobility Management Entity}
\acrodef{amf}[AMF]{Access and Mobility Management function}
\acrodef{mvno}[MVNO]{Mobile Virtual Network Operations}
\acrodef{mno}[MNO]{Mobile Network Operator}

\newcommand{\eg}{\textit{e.g.,}\xspace}
\newcommand{\etc}{\textit{etc.}\xspace}
\newcommand{\ie}{\textit{i.e.,}\xspace}

\newcommand{\fig}{Fig.~}
\newcommand{\tab}{Table~}

\newcommand{\simpletitle}[1]{\noindent\textbf{#1}\xspace}

\begin{document}
\bstctlcite{IEEEexample:BSTcontrol}
\title{Spectrum \& RAN Sharing: A Measurement-based Case Study of  Commercial 5G Networks in Spain\\

\thanks{}%
}
\author{
\IEEEauthorblockN{Rostand A. K. Fezeu\IEEEauthorrefmark{1}, Lilian C. Freitas\IEEEauthorrefmark{1}, Eman Ramadan\IEEEauthorrefmark{1}, 
Jason Carpenter\IEEEauthorrefmark{1}, 
\\ Claudio Fiandrino\IEEEauthorrefmark{2}, Joerg Widmer\IEEEauthorrefmark{2} and Zhi-Li Zhang\IEEEauthorrefmark{1}}
\IEEEauthorblockA{
\textit{\IEEEauthorrefmark{1}University of Minnesota} -- Minneapolis, USA | 
\IEEEauthorrefmark{2}\textit{IMDEA Networks Institute} -- Madrid, Spain\\
\{fezeu001, ldefreit, eman, carpe415\}@umn.edu, zhzhang@cs.umn.edu, \{claudio.fiandrino, joerg.widmer\}@imdea.org}}

\maketitle

\begin{abstract}
\ac{ran} sharing, which often also includes spectrum sharing, is a strategic cooperative agreement among two or more mobile operators, where one operator may use another's RAN infrastructure to provide mobile services to its users.  By mutually sharing physical sites, radio elements, licensed spectrum and other parts of the RAN infrastructure, 
participating operators can significantly reduce the capital (and operational) expenditure in deploying and operating cellular networks, while accelerating coverage expansion -- thereby addressing the spectrum scarcity and infrastructure cost challenges in the 5G era and beyond.  While the economic benefits of RAN sharing are well understood, the impact of such resource pooling on user-perceived performance remains underexplored, especially in real-world commercial deployments. We present, to the best of our knowledge, the first empirical measurement study of commercial 5G spectrum and \ac{ran} sharing. Our measurement study is unique in that, beyond identifying real-world instances of shared 5G spectrum and \ac{ran} deployment ``in the wild'', we also analyze users' perceived performance and its implication on \ac{qoe}. Our study provides critical insights into resource management (\ie pooling) and spectrum efficiency, offering a blueprint (and implications) for network evolution in 5G, 6G and beyond. 

\end{abstract}

\begin{IEEEkeywords}
5G NR, RAN sharing, spectrum sharing,  measurements, QoE.
\end{IEEEkeywords}

\section{Introduction}
\label{s:intro}
Radio spectrum and radio access network (RAN) infrastructure are two major expenses in cellular network deployment and operations. According to~\cite{substackMostExpensive}, since 2000, telecommunications operators have paid approximately \$1~trillion for spectrum access, with up to 45\% upfront in primary license awards, \$50–100 billion in clearing and refarming, and remaining on recurring spectrum license fees. Likewise, based on the 2024 key industry statistics~\cite{wiaWirelessInfrastructure}, Wireless Infrastructure Association (WIA) estimates that the U.S. cellular industry spent more than \$10.8 billion  in deploying new and upgraded networks in 2024 alone, of which \$8.4 billion was spent on RAN. 

To partly lower  and share the cost burdens, 3GPP has introduced the concept of  \emph{network sharing} in 3GPP Specification TS~23.251~\cite{ranSharing}, which allows multiple network operators to share radio access elements (\emph{RAN sharing}), which may also include the sharing of the radio resources themselves (\emph{spectrum sharing}). The main type of network sharing is Multi Operator Core Network (MOCN), in which multiple mobile operators with their separate dedicated core networks share some radio access elements (e.g., a base station) at certain locations.
RAN sharing and spectrum sharing (as part of RAN sharing) are in particular deemed as important mechanisms in 5G deployment to share spectrum, improve coverage, and reduce costs~\cite{network-sharing-5G}. In 2024, indirect network sharing (INS) was introduced in 3GPP 5G-Advanced Release 20 as the next step in network sharing evolution~\cite{TS23501} (see \S\ref{s:background} for more details). 

\begin{figure}[t]
    \centering
\includegraphics[scale=1, keepaspectratio]{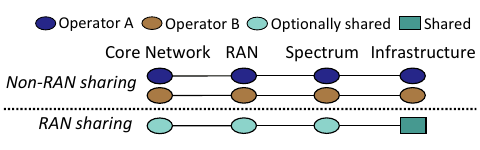}%
     \vspace{-1em}%
    \caption{Shared and non-shared network elements in RAN Sharing.}%
    \label{fig:sharing}%
      \vspace{-1em}
\end{figure}

At the first glance, network sharing may sound similar to the situation of \ac{mvno}~\cite{wiki-MVNO}, where a \ac{mvno} leases
network access from an MNO (Mobile Network Operator) to offer/resell mobile services without owning the physical infrastructure -- hence in a sense the \ac{mvno} ``shares'' the physical infrastructure of the MNO. Network sharing, as defined by 3GPP, however, refers specifically to the cases where multiple MNOs share parts of their network infrastructure (see \fig\ref{fig:sharing} for an example). Hence, it also differs from \emph{roaming},  where a ``foreign'' (visited) MNO provides temporary mobile access to a visiting user from another MNO. In a nutshell,  the goal of network sharing is to allow multiple MNOs to share part of their network infrastructure to reduce costs and improve coverage.  

With the rollout of 5G networks (and enormous associated costs) since 2019, several RAN sharing agreements between MNOs have been publicly announced in multiple countries (see, e.g.,~\cite{vodafoneOrangeSpain2019,Japan2024, China2025_5GA}). Several white papers/research papers~\cite{gsma2023_China,Bourreau_2020, Ivaldi2021RANSharing,koutroumpis2023impact} have also been published that analyze the potential economic benefits of RAN sharing.  For instance, SoftBank and KDDI in Japan both claimed~\cite{Japan2024} that by sharing 4G base station infrastructures, they are able to cut capital expenditure by about \$288 million for each company, with (part of the) cost savings often passed down to customers through lower service prices.  Beyond just cost savings, RAN sharing can also bring potential performance benefits for users, as suggested by several  simulation-based studies~\cite{Lin2025, Zhao2021}.   While the notion of RAN sharing has been introduced by 3GPP for some time and 5G networks have been rapidly rolled out since 2019, there have been no \emph{publicly available} research studies that identify the existence of RAN sharing in \emph{real world 5G deployment}, not to mention quantifying the performance impacts of RAN sharing on mobile applications from a user perspective. This paper aims to fill this gap. 

We have conducted an extensive measurement study of commercial 5G network deployment in Spain. Our study provides the first measurement-based, quantitative evidence of 5G network sharing ``in the wild'' where two operators, Vodafone and Orange, not only share the RAN infrastructure but also spectrum (an n78 channel owned by Vodafone) in Spain. This finding is further backed up by the Spanish government spectrum auction data as well as the public announcement of a network-sharing agreement between the two operators~\cite{vodafoneOrangeSpain2019}. With the discovery of RAN (and spectrum) sharing in a real-world deployment scenario,  we further investigate the following questions: 1) How do commercial spectrum and \ac{ran} sharing affect resource management in the 5G era?  2) With RAN sharing, do both operators benefit equally from the performance perspective? 3) Does an operator provide similar performance to its users with or without RAN sharing? And, 4) Does resource pooling afforded by RAN \& spectrum sharing translate to improve \ac{qoe} for the users?  We summarize the main contributions of our measurement-based inquiry of RAN and spectrum sharing below.

\noindent$\bullet$ We advance commercial 5G understanding and conduct -- to the best of our knowledge, the first large-scale, comprehensive, and comparative measurement study of commercial 5G spectrum and \ac{ran} sharing in Spain, filling an important gap in the existing 5G measurement literature. 

\noindent$\bullet$ Through our detailed measurement analysis, we evaluate and study the performance characteristics of RAN sharing and non-RAN sharing, further quantifying its direct effects on application \ac{qoe}. 

\noindent$\bullet$ Our measurement provides insights into deployment strategies, studying the intricacies of resource pooling when RAN sharing, and its implications for 5G evolution.

While our study presents a \emph{real-world} case study of RAN and spectrum sharing in Spain, we believe that both the measurement methodology we have developed as well as the findings and insights we have obtained will have wide applicability in other countries and scenarios. Mobile industry and the 3GPP standard organization are now in the midst of 6G radio and network architecture and technology specifications.  With increasing demands and costs for spectrum~\cite{GSMA-Global-Spectrum-Pricing}, it is widely believed that (dynamic) spectrum sharing will be a key feature among many others that are under study. The costs of deploying new 6G radio elements or upgrading the existing 5G  RAN infrastructure to future 6G standards, however, are a major hurdle for \ac{mno}s to fully embrace 6G~\cite{when-operators-want-operators}. Spectrum and RAN sharing -- and, more broadly, network sharing -- will therefore play a critical role in reducing costs, expanding coverage, improving performance, and enhancing user experience, thereby benefiting both operators and end users.

\section{Background \& Motivation}
\label{s:background}

In this section, we present the technical background and introduce key concepts and terminology related to RAN sharing, including both passive and active RAN sharing architectures. We then discuss the motivation and key research questions that guide our work.

\begin{table*}[th]
\centering
\begin{minipage}{0.97\textwidth}
\caption{Comparison of RAN sharing architectures.}%
\label{tab:rancomparison}%
\centering%
\setlength{\tabcolsep}{6pt}
\setlength{\extrarowheight}{10pt}
\resizebox{0.95\width}{!}{%
\centering
\begin{tabular}{c c c c}
\hline
\textbf{Type of Sharing }
& \textbf{Scenario}
& \textbf{What is Shared}
& \textbf{Key Characteristics}
\\ \bottomrule 
Passive RAN Sharing
&  
& \makecell{Civil infrastructure only, \ie Sites, towers,\\  masts, power, cooling, backhaul \etc} 
& \makecell{Cost-driven model; no sharing of radio \\ equipment, spectrum, or RAN functions. }

\\ \bottomrule 
\multirow{5}{2.2cm}{\makecell{Active RAN Sharing}}
&  MORAN
& \makecell{Infrastructure + RAN nodes (\eg NodeB, \\eNB, gNB), antennas, radio equipment}
& \makecell{Active RAN sharing while retaining \\independent spectrum ownership. }
\\ \cline{2-4}

&  MOCN
& \makecell{Infrastructure + RAN + Spectrum resources}
& \makecell{Widely deployed active sharing model. \\Improves spectrum utilization \& cost efficiency. }

\\ \cline{2-4}

&  GWCN
& \makecell{Infrastructure + RAN + spectrum resources \\+ part (or all) of core network}
& \makecell{Greater cost savings but reduces\\ flexibility and roaming support.}
\\ \cline{2-4}

&  INS
& \makecell{Infrastructure + RAN \\ Traffic routed via a ``master'' operator core network}
& \makecell{Operational arrangement similar to \\roaming but within a sharing agreement.}
\\ \bottomrule 

\end{tabular}
}
\end{minipage}
\end{table*}

\subsection{Spectrum and RAN Sharing Background}
\label{ss:bck}

In the traditional model of single ownership, an operator is entirely responsible for: (i)~acquiring the spectrum, typically through regulated auction or other mergers (e.g., merging companies, short-term or geographic leasing) -- spectrum acquisition is one of the biggest expenses in cellular deployment. (ii)~deploying, operating, and maintaining its own infrastructure (i.e., sites,  towers, backhaul), spectrum, RAN, and core network, incurring substantial capital and operational expenditure. (iii)~ provisioning, managing services and executing subscriber billing, charging and accounting functions~\cite{Cano2020}. In the spectrum and RAN sharing model, two or more operators jointly deploy, operate, and maintain the network\footnote{3GPP allows RAN sharing among up to six operators~\cite{TS32130}.}. Operators may share different parts of the network, including infrastructure, spectrum, RAN, and even part of the core network, while remaining independent~\cite{TS32130}. For the network subscriber, the services appear normal without interruptions. 

The motivation for Spectrum/RAN sharing is to reduce cost while enhancing network capacity. RAN sharing was widely adopted through successful transitions between mobile network generations (\ie 2G$\rightarrow$3G and 3G$\rightarrow$ 4G), only to reduce the investment cost required for these technology migrations. In the case of 5G with ultra-high throughput requirements, RAN sharing is not just to reduce costs, but primarily to address spectrum scarcity. Several RAN sharing architectures have been discussed in industry and have been standardized within 3GPP~\cite{TS32130}. Based on which network elements the operators agree to share, there are two main types of sharing: passive and active RAN sharing. \tab~\ref{tab:rancomparison} summarizes the main RAN sharing architectures.

\simpletitle{Passive RAN sharing:} This is also known as infrastructure sharing according to 3GPP TS~32.130~\cite{TS32130}. In this type of sharing, only the physical site and civil infrastructure (like the towers, rooftop space, equipment shelters, backup power (batteries, generators), cooling/air conditioning, grounding, and cabling trays) are shared. The operators retain independent spectrum, RAN, and core network infrastructures.

\simpletitle{Active RAN sharing:} Here, different network elements are shared and can be organized as follows:

\noindent \textit{1)~Multi-Operator Radio Access Network (MORAN):} In MORAN, the infrastructure and RAN are shared. Spectrum and core network resources are not pooled. As a result, operators with large spectrum will benefit from high capacity~\cite{Larsen2023}. This is an industry RAN sharing, not defined by 3GPP. 

\noindent \textit{2)~Multiple Operator Core Network (MOCN):}  Here, not only the infrastructure and RAN are shared, but also the spectrum. However, operators maintain separate core networks~\cite{TS23251, TS23501}. In this RAN sharing scenario, spectrum resources scheduling is done fairly across users belonging to different operator~\cite{Larsen2023}. Unlike MORAN, this type of RAN sharing is defined by 3GPP.

\noindent \textit{3)~Gateway Core Network (GWCN):} In this 3GPP sharing architecture, in addition to the infrastructure, spectrum, and the RAN, part or all of the core network is also shared~\cite{TS23251}. This provides greater cost savings at the expense of reduce flexibility, especially in mobility scenarios which involves several control plane network functions.

\noindent \textit{4)~Indirect Network Sharing (INS):} Infrastructure and RAN are shared. Participating operators do not have direct connections to the shared RAN. Communication between the shared RAN and the core network of the participating operators is routed via the core network of the master operator, similar to roaming with joint RAN operations.

\subsection{5G Background and Motivation}
\label{ss:background_Motivation}

\simpletitle{Brief 5G Background:}In 5G, when a \ac{ue} wants to send/receive any voice/data via the cellular network, it must first establish a connection with the base station. The \ac{ue} uses the subscriber network channel spectrum, which determines the available \ac{prb} during scheduling -- a resource block is the basic unit of radio resource allocation~\cite{ross2024midband}. To establish connection with the core network, the \ac{ue} must also undergo several control plane operators~\cite{ross_cp}.  A key control plane operation that is triggered is the Radio Resource Control (RRC) procedure. The RRC Connection procedure is initiated to transition the UE from an \textit{RRC\_IDLE} to \textit{RRC\_CONNECTED} state. A UE must be in \textit{RRC\_CONNECTED} state before it can send/receive data on the network. During the RRC procedure, another key  connection establishment procedure is triggered -- The \ac{pdn} in 4G LTE or \ac{pdu} connection procedure in 5G. The PDN/PDU connection procedures provide \ac{e2e} user/data plane connectivity between the UE via the RAN through the core network. Therefore, depending on which components are shared, the performance experienced by a user (both in terms of latency and throughput) may differ, as we later show in~\S\ref{s:results}. 

\simpletitle{Motivation:}Consider the case in which two operators, in addition to sharing the infrastructure and spectrum, (i)~\textit{ also share the RAN} or (ii)~\textit{share part or all of the core network}. In case~(i), users’ voice/data traffic must be routed to the respective core networks, or in case~(ii), traffic is routed to the shared core network. Questions arise around which operator manages how a user connects to the RAN during mobility. What is the impact of sharing part or all of the RAN and/or core on \acp{ho} -- The key mobility management procedure involved when the UE's voice/data must be redirected to a new base station (BS) with better signal quality~\cite{ahmad_sigcomm}. This raises important  ``yet unanswered questions'': (1) \textit{What is the performance of control plane latency of RAN sharing compared to non-RAN sharing?} (2) \textit{Does resource pooling when RAN sharing provide improve data plane performance for users?} (3) \textit{How does it impact the user \ac{qoe}?}
\section{Spectrum Acquisition in Spain and Measurement Methodology}
\label{s:method}
 
In this section, we start by discussing spectrum acquisition in Spain, contrasting spectrum auction outcomes based on publicly available information with those observed in our measurement data. We highlight the intricacies and complexities of other operator agreements and spectrum exchanges that occur behind publicly available information. We show how we used our measurement data to identify that Orange and Vodafone in Spain are involved in active RAN sharing. We conclude this section with a detailed discussion of our data-collection methodology.

\subsection{Spectrum Acquisition in Spain}
\label{s:spectrumAcquisition}

Although 3GPP specifies ``allowable'' channel bandwidths for each 5G band\footnote{For example, for the n78 band, the specified channel bandwidths are 10, 15, 20, 25, 30, 40, 50, 60, 70, 80, 90, and 100 MHz.}, the bandwidth deployed in practice is largely determined by the spectrum acquired by each operator through auctions, mergers, or other acquisition mechanisms. This distinction has direct implications for RAN sharing, as the effective channel bandwidth depends on both spectrum holdings and the sharing arrangements between participating operators. It is important to note that spectrum acquisition is one of the largest cost components in 5G deployments. In this work, we use the Spanish telecommunications market as a case study to analyze how regulatory outcomes translate into operational network deployments.\\
\simpletitle{Auction Outcomes:} From public sources, the Spanish spectrum authority distributed the n78 band (3.4~--~3.8~GHz) with three separate auctions. (i)~2016 Auction (3.4~--~3.6~GHz)~\cite{Spain-n78-auction-1}: After this auction, MasMovil (now part of Yogio Spain) secured 80 MHz in two non-contiguous 40~MHz channel/spectrum blocks; 3.4~--~3.44~GHz and 3.5~--~3.54~GHz. Telefonica and Orange each acquired 40~MHz, split into 20~MHz segments. (ii)~2018 Auction (3.6~--~3.8~GHz)~\cite{Spain-n78-auction-2}: In this phase, Vodafone Spain spent 198.1 million euros to acquire eighteen 5~MHz blocks for a total of 90~MHz and Orange Spain spent 132.2 million euros and obtained 60 MHz (twelve 5~MHz) channels, while Telefonica spent 107.4 million euros to purchased a 50~MHz (ten 5~MHz) channel bandwidth. (iii)~In 2021~\cite{Spain-n78-auction-3}, Telefonica and Orange each further bid and purchased two 10~MHz channels\footnote{In the first auction in 2016, the Spanish spectrum authority reserved two 20 MHz channels in the 3.4~--~3.6~GHz range for military radio service. Telefonica's 20 MHz acquisition in the third auction is likely those reserved, according to~\cite{Spectrum-tracker-spain}.}. The remaining 3.48~--~3.5~GHz channels were acquired by Yogio Spain. As a summary, after the auction, in the n78 band, Orange owned 110~MHz, Telefonica owned 100~MHz, Vodafone acquired 90~MHz, and Yogio purchased 80~MHz. \\
\simpletitle{``In the Wild'' Observation:}From the above auction outcomes, Orange and Telefonica in Spain both own non-contiguous, scattered channel bandwidths, which require Carrier Aggregation (CA) to increase data rates~\cite{weiye2024ca}. However, measurement data reveals that Orange is now operating a contiguous 100 MHz channel (3.6~--~3.71~GHz) -- a block that overlaps with Telefonica's 2018 acquisition. This observation suggests a post-auction spectrum swap between Orange and Telefonica to make their spectrum contiguous. We later confirmed this by ~\cite{Spectrum-tracker-spain}, which indicates that Telefonica now controls a contiguous 100~MHz (3.5~--~3.6~GHz) channel. \\
\simpletitle{Spectrum and Active RAN Sharing:}Interestingly, based on our measurement, both Orange and Vodafone use the exact same 90 MHz channel (3.71~--~3.8 GHz). While Vodafone remains the sole legal owner of this specific block~\cite{Spectrum-tracker-spain}, Orange Spain utilizes it via a spectrum and RAN infrastructure sharing (and part of the core network) agreement with Vodafone Spain~\cite{Bourreau_2020}. As a result, Orange is able to extend its 5G footprint by using Vodafone's spectrum and RAN infrastructure in specific geographic areas. This active spectrum and RAN sharing represents a commercial deployment for the RAN sharing specified by 3GPP~\cite{ranSharing}.
Next, we present a detailed discussion of our measurement methodology and data collection process.

\subsection{Measurement Methodology \& Data Collection}
\label{ss:method}

\begin{figure*}[t!]
\centering
\begin{minipage}[c]{0.3\textwidth}
\centering
\includegraphics[width=0.98\textwidth, height=4.2cm, ]{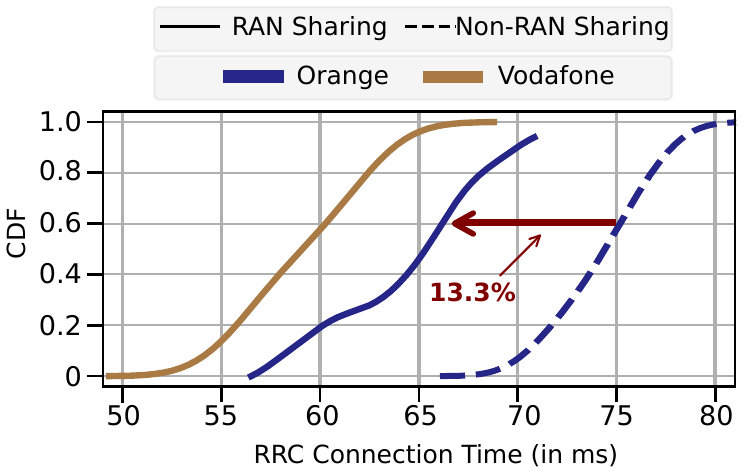}%
    \vspace{-1em}
    \caption{RRC connection time  with and without RAN Sharing.}%
    \label{fig:rrc_connection}
\end{minipage}%
\hfill%
\begin{minipage}[c]{0.3\textwidth}
\centering
    \includegraphics[width=0.98\textwidth, height=4.2cm,]{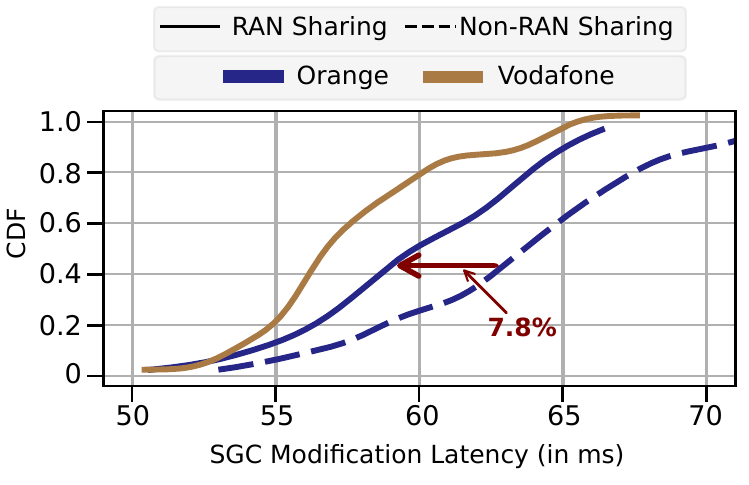}%
    \vspace{-1em}
    \caption{SGC Modification latency with and without RAN Sharing.}%
    \label{fig:sgc_latency}
\end{minipage}%
\hfill
\begin{minipage}[c]{0.3\textwidth}
\centering%
    \includegraphics[width=0.98\textwidth, height=4.2cm,]{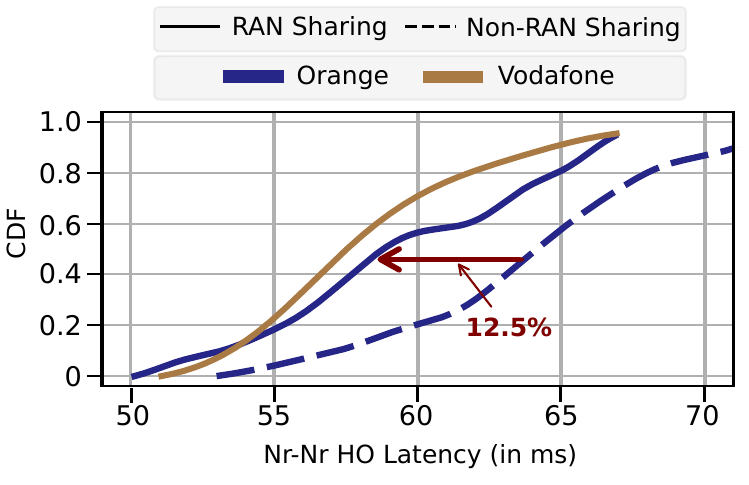}%
    \vspace{-1em}
    \caption{NR-NR HO Latency with and without RAN Sharing.}%
    \label{fig:ho_latency}
\end{minipage}%
\vspace*{-4ex}%
\end{figure*}

Our RAN Sharing study and analysis is made possible by carefully orchestrating data collection that focuses on three dimensions: (1)~\textit{5G operators}, (2)~\textit{measurement platform and applications}, and (3)~\textit{orchestration of data collection.}\footnote{A detailed discussion of our measurement campaign is outlined in our recently published paper~\cite{ross2024midband}. Refer to~\cite{roaming-paper} for a focused measurement discussion of our Roaming study. Here, we provide a discussion of our measurement campaign focused on RAN sharing in Spain.} 

\simpletitle{5G Operators:} We relied on public information available on the Internet to identify countries with operators that have come to a spectrum/RAN sharing agreement. We identified Orange and Vodafone in Spain~\cite{Bourreau_2020}, and also confirmed this using our measurement data as mentioned earlier. We next relied on several related works like~\cite{fiandrino2022uncovering, 10.1145/3517745.3561465}, and platforms like nperf~\cite{nperf} and Ookla speedtest~\cite{ookla_speedtest} to identify cities in Spain where Orange and Vodafone have deployed 5G services. Both Orange and Vodafone deploy their 5G services in 5G mid-band, \ie 3.3--3.8 GHz, in band n78. We next purchased four contract SIM cards (two per operator) to avoid performance throttling. 

\simpletitle{Measurement Platform and Applications Tested:}Since our goal is to study spectrum/RAN sharing performance from both the network (\ie PHY and MAC layers) and user (application) perspective, we carefully designed a comprehensive measurement platform. We selected a diverse set of testing servers and applications. We leveraged Ookla speedtest servers deployed within/close to Orange and Vodafone's networks in Madrid, Seville, Barcelona, Bilbao, and Valencia to measure and compare the \ac{e2e} latency, ``raw'' downlink (DL) and uplink (UL) throughput under RAN sharing and non-RAN sharing. We additionally conducted \texttt{iPerf}, ping, and traceroute experiments using cloud servers. We deployed cloud servers in Google Cloud Platform (GCP)~\cite{gcp}, Microsoft Azure Cloud~\cite{acp}, and Amazon AWS Cloud~\cite{AWS}, the three major cloud service providers. To study the impact of spectrum/RAN sharing on users' application \ac{qoe}, we also design and conduct video streaming experiments. 

Unlike in-lab experiments conducted in a controlled environment, conducting experiments in the wild poses several challenges. Therefore, we deployed automated custom scripts to ensure reliable data collection (to the extent possible). Since the phone's capabilities have been shown to affect a user's performance~\cite{weiye2024ca, ross2024midband}, we use four phones with the same model, Samsung Galaxy S21 Ultra, for all our experiments. To extract measurements across the 5G network-- \ie 5G New Radio (NR) -- we use Accuver XCAL~\cite{xcal}, a professional cellular measurement tool. The smartphones are connected via USB-C to a Windows laptop running XCAL. XCAL supports up to six devices simultaneously and enables concurrent data collection across the full 5G NR protocol stack, with data extracted directly from the 5G modem chipsets. We conducted both stationary and drive-test experiments.

\simpletitle{Data Collection: } Using our measurement platform, we conducted an extensive 5G RAN sharing measurement campaign. For data collection, we performed drive tests across Madrid, running downlink \texttt{iPerf} experiments to identify 5G areas where Orange and Vodafone operate using shared 5G spectrum under their agreement~\cite{Bourreau_2020}. We identified two such areas around Madrid. Subject to financial constraints, we conducted both stationary and drive-test experiments and collected detailed measurements for RAN sharing and non-RAN sharing scenarios across Madrid, Spain, on different weekdays and weekends. In summary, we collected more than 1,060 minutes of non-RAN sharing measurements and over 1,100 minutes of RAN sharing measurements across 5G networks, totaling over 1.2~TB of data.

\section{ Performance Analysis of RAN Sharing}
\label{s:results}

In this section, we use our measurement data to compare performance under RAN-sharing and non-RAN-sharing scenarios. Specifically, we compare the control-plane latency and throughput for RAN sharing and non-RAN sharing between Orange and Vodafone in Spain. We conclude by showing how the “raw” throughput performance affects users, an issue that, to the best of our knowledge, has not yet been studied within the scope of 5G. 

\subsection{RAN Sharing Impact on Control Plane Latency}
\label{ss:latencyCP}

We quantitatively study the control plane latency with and without RAN sharing. Recall from \S\ref{ss:background_Motivation} that, when a \ac{ue} needs to send/receive voice/data over the cellular network, it must transition into the \textit{RRC\_CONNECTED} state. The process involves establishing a data \ac{pdn}/\ac{pdu} session to the operator's core network. Here we compare the latency of three key control plane procedures under RAN sharing and non-RAN sharing: \textbf{(i)~\textit{ RRC Connection Time: }}The time elapsed for a \ac{ue} to transition from an \textit{RRC\_IDLE} state to an \textit{RRC\_CONNECTED} state to send/receive voice/data. This procedure includes the time required to establish a \ac{pdn}/\ac{pdu} connection to the core. \textbf{(ii)~\textit{Secondary Cell Group (SCG) Modification Latency: }}The delay incurred when changing the secondary base station group connected to the UE while maintaining the primary cell connection. \textbf{(iii)~\textit{NR-NR Handover (HO) Latency: }}The delay to redirect a UE's voice and data to a new base station with better signal quality.  

We study the control-plane latency performance when an Orange UE connects to its home network, Orange -- \ie non-RAN sharing, and when the same Orange UE connects to Vodafone -- \ie RAN sharing. \fig\ref{fig:rrc_connection} shows the \textit{RRC Connection Time} for Orange subscriber under RAN sharing and non-RAN sharing. We observe that the \textit{RRC Connection Time} drops (\ie improves) by 13.3\% with RAN sharing. We also find that the modification of secondary cell groups (SCG latency) using RAN sharing also drops (improves) by 7.8\%, as illustrated in \fig\ref{fig:sgc_latency}. These findings indicate that non-shared cells are also added to a UE cell group.  Additionally, we quantify the \textit{NR–NR HO latency} of the Orange UE while driving using RAN sharing and non-RAN sharing, as shown in \fig\ref{fig:ho_latency}. We observe a 12.5\% drop (\ie improvement) in control-plane latency for Orange under RAN sharing compared to non-RAN sharing. 

When we compare a Vodafone user with an Orange user, under RAN sharing, we observe that the \textit{RRC Connection Time} and \textit{SCG latency} are 8.03\% and 4.71\% lower (\ie better) for Vodafone users (see \fig\ref{fig:rrc_connection} and \fig\ref{fig:sgc_latency}). However, these differences are smaller when we compare the Vodafone user with the baseline non-RAN sharing Orange performance. Although not explicitly shown in the figures, we observe that Vodafone users exhibit the same control-plane latency performance during RAN sharing as in the non–RAN sharing case. In other words, Vodafone users maintain consistent control-plane latency \ac{qoe}, even when actively sharing the RAN with Orange users. 

\begin{figure}[b!]
    \centering
    \vspace{-1em}%
    \includegraphics[scale=0.6, keepaspectratio]{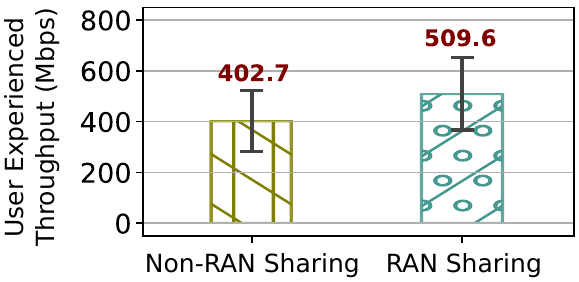}%
    \vspace{-1em}%
    \caption{Comparison of Orange Subscriber UE throughput under non–RAN sharing and RAN sharing scenarios.}%
    \label{fig:UE_TputDL}%
\end{figure}

\begin{figure*}[t!]
\centering

\begin{minipage}[c]{0.45\textwidth}
\centering
    \includegraphics[scale=0.5, keepaspectratio]{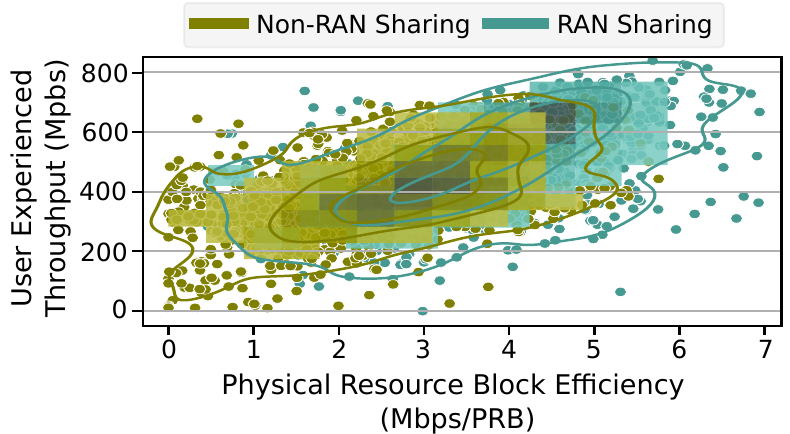}%
     \vspace{-1em}%
    \caption{[Orange] PRB efficiency and UE throughput under RAN sharing and non–RAN sharing scenarios.}%
    \label{fig:RB-Efficiency}%
\end{minipage}%
\hspace{2ex}
\begin{minipage}[c]{0.45\textwidth}
\centering%
    \includegraphics[scale=0.6, keepaspectratio]{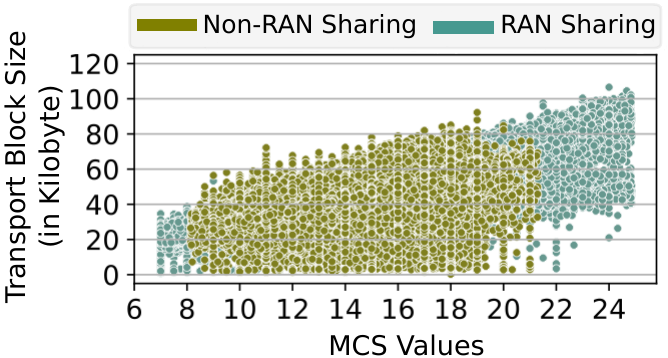}%
    \vspace{-1em}%
    \caption{[Orange] Transport Block Size (TBs) and MCS under RAN sharing and non–RAN sharing scenarios.}%
    \label{fig:mcs}%
\end{minipage}%
\vspace*{-4ex}%
\end{figure*}
These results imply that Orange and Vodafone in Spain also share part of the core network. We further confirm this by extracting the \ac{mme} code, which is 232 in our study. The \ac{mme} code includes the \ac{amf} identifier (a core network function) to which the UE is connected. Since the \ac{mme} code is identical under both RAN sharing and non-RAN sharing scenarios, Orange and Vodafone are likely sharing part of the core network. We note that \ac{amf} sharing may introduce potential security concerns, which remain to be investigated. The reduced control plane latency resulting from sharing part of the core network is consistent with the behavior expected under GWCN active RAN sharing. Since our analysis is based on UE-side measurements, this observation provides indirect evidence rather than definitive confirmation of the underlying network architecture. Overall, our results highlight the benefits of resource pooling under RAN sharing, particularly in reducing control-plane delay. Next, we quantitatively analyze data-plane performance.

\subsection{RAN Sharing Impact on Throughput}
\label{ss:latency_dataplan}

To understand how RAN sharing impacts user performance, we quantify the throughput under RAN sharing and non-RAN sharing scenarios. \fig~\ref{fig:UE_TputDL} shows the \ac{ue} downlink throughput (y-axis) for the non–RAN sharing and RAN sharing cases. We observe a 26.54\% increase in throughput, from 402.7~Mbps to 509.6~Mbps, when the UE operates under RAN sharing compared to non–RAN sharing.

To further analyze resource utilization, we quantify the \ac{prb} efficiency -- a measure of how effectively the network utilizes \acp{prb} -- under RAN sharing and non–RAN sharing scenarios. \fig~\ref{fig:RB-Efficiency} shows a scatter plot of the \ac{prb} efficiency (Mbps/\ac{prb}) on the x-axis and \ac{ue} throughput on the y-axis. We observe that the network achieves higher \ac{prb} efficiency under RAN sharing. We suspect that this gain is primarily due to resource pooling, which is known to optimize \ac{prb} allocation~\cite{qadir2016resource}. For instance, as shown in~\fig~\ref{fig:mcs}, when pooling resources, operators can implement and utilize higher Modulation and Coding Scheme (MCS) values (x-axis), which in turn increases the amount of data sent on the wireless link (\ie Transport Blocks Size) (y-axis), and consequently increases the throughput. Additionally, in RAN sharing deployments, operators can mitigate redundant control-plane messages and procedures and optimize the \ac{ho} process -- as also demonstrated in \fig\ref{fig:ho_latency} -- thereby freeing up additional resource blocks. This allows more user data to be multiplexed, resulting in higher \ac{prb} efficiency and, consequently, increased throughput. Such performance gains highlight the benefits of RAN sharing in mobile networks, particularly for 5G/NextG applications that require ultra-high throughput.

\subsection{RAN Sharing Impact on Application QoE}
\label{ss:app}

Our objective is to quantitatively evaluate the impact of RAN sharing on user-perceived \ac{qoe}. We conducted video streaming experiments in both RAN sharing and non–RAN sharing scenarios.

\simpletitle{Experiment Approach:}In these experiments, a \ac{ue} streams a video by downloading video chunks from a remote server located in the same geographical area as the \ac{ue} (both are in Madrid, Spain). The video is divided into fixed-length chunks and encoded with different quality levels. We used FFmpeg with libx264 to encode a 210-second video into seven quality levels with different bitrates at 30 frames per second (fps) and a chunk length of 4 seconds, which is the recommended chunk duration for Adaptive Bitrate (ABR) streaming~\cite{Lederer_2020}. The seven quality levels correspond to approximate bitrates of 30~Mbps, 60~Mbps, 75~Mbps, 200~Mbps, 400~Mbps, 600~Mbps, and 750~Mbps. As throughput varies over time in both RAN sharing and non–RAN sharing scenarios, the ABR algorithms dynamically adapt the requested video chunk quality to improve \ac{qoe}. For our implementation, we used DASH.js~\cite{dash}, a popular open-source video streaming framework, and deployed two ABR algorithms: BOLA~\cite{bola} and a dynamic bitrate algorithm. We used a custom HTML-based player to conduct trace-based emulation under both RAN sharing and non–RAN sharing conditions.
For our analysis, we normalized the stall time percentage (\ie the fraction of time elapsed while waiting for a video chunk to be played) and the average bitrate of the delivered video chunks.

\begin{figure}[h!]
    \centering%
    \vspace{-1em}
    \includegraphics[scale=0.8, keepaspectratio]{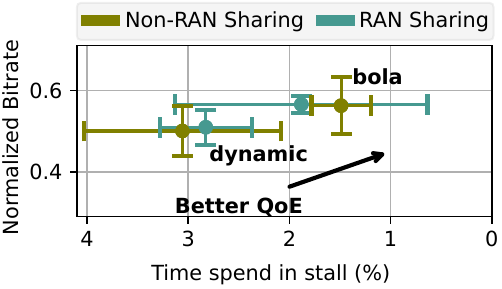}%
    \vspace{-1em}
    \caption{[Orange] QoE Performance under RAN Sharing and Non-RAN sharing.}%
    \label{fig:abrCarrier}%
    \vspace{-1em}
\end{figure}

\fig~\ref{fig:abrCarrier} presents the normalized average bitrate (y-axis) and stall time percentage (x-axis) for video streaming in the RAN sharing and non–RAN sharing scenario.
We observe a slight increase in bitrate in the RAN sharing scenario, regardless of the ABR algorithm, which is consistent with the slight increase in throughput observed under RAN sharing earlier. These results suggest that current ABR algorithms play a limited role in improving \ac{ue} \ac{qoe} in the RAN sharing scenario. A more detailed analysis of the dynamics of specific ABR algorithms during streaming, particularly the differences in stall time performance between RAN sharing and non–RAN sharing scenarios, is left for future work. Nonetheless, our analysis provides insight into how RAN sharing can improve \ac{qoe}.
\section{Related Work}
\label{s:related_works}

We organize the related work on RAN sharing into three categories: (i)~studies on the theoretical economic and technical potential of RAN sharing, (ii)~other commercially deployed ways in which operators collaborate, such as \ac{mvno} and Roaming, which are often discussed in the literature and sometimes confused with RAN sharing, and (iii)~academic experimental studies on RAN sharing. We conclude this section by articulating the research gap our work aims to address and clearly discussing how it contributes to advancing the understanding of commercial 5G deployments.

\subsection{Economic and Technical Potential of RAN Sharing}

Several RAN sharing deployments have been implemented by commercial mobile networks worldwide~\cite{vodafoneOrangeSpain2019,gsma2023_China, Japan2024}. 
Studies conducted in~\cite{Bourreau_2020, Ivaldi2021RANSharing, koutroumpis2023impact} show that active RAN sharing significantly enhances both network quality and market competition. Koutroumpis \textit{et al.}~\cite{koutroumpis2023impact} analyzed the economic and technical outcomes of RAN sharing using an extensive real-world dataset covering 140 mobile operators across 29 European countries over a 20-year period (2000–2019). These studies indicate that sharing leads to faster technology rollouts (4G/5G), broader coverage, and increased download speeds. Furthermore, these technical gains do not come at the cost of competition; instead, the operational savings are often passed down to consumers, resulting in lower service prices. While passive RAN sharing provides cost relief to operators, active RAN sharing is the clear driver for superior technical performance, enabling smaller operators to provide service quality comparable to that of larger operators.

\subsection{Mobile Virtual Network Operators \& Roaming}

In addition to RAN sharing, operators use other collaborative service models to reduce costs, such as \textit{roaming} and \textit{Mobile Virtual Network Operator (MVNO)}. 
\textit{Roaming} is a widely adopted service-level agreement that ensures service continuity when users are outside their home network's coverage.  Visited network provides radio access; authentication and billing are anchored in the home network. Most mobile operators maintain such agreements, aiming to expand network coverage; however, most roaming agreements impose limits on the maximum data volume and service speed available to roaming users, which can negatively affect the resulting \ac{qoe}~\cite{Mahmood2025}.
\textit{\ac{mvno}} is a commercial/wholesale agreement that enables a mobile network to operate without deploying full infrastructure, but instead leases network capacity from a host operator to provide retail services under its own brand. An MVNO is identified by its operation without ownership of licensed spectrum, obtaining access through agreements with a licensed operator~\cite{AlcalaMarin2024}. 
Unlike RAN sharing architectures, roaming and MVNO arrangements involve collaboration at the service and business levels, while the network infrastructure remains owned and managed independently by each operator.

\subsection{Academic RAN Sharing Studies}

Most existing studies on RAN sharing rely on simulation data or experimental laboratory networks~\cite{Lin2025, Zhao2021}, while only a limited number examine performance in commercial (live) mobile networks from a user perspective.
Türk and Zeydan~\cite{Turk2021} show that active RAN sharing can effectively enhance network capacity and user experience without degrading core network performance. They conducted an experimental RAN sharing trial on live 4G networks in Turkey. They enable RAN sharing at operational sites and systematically compare performance metrics before and after the sharing configuration to quantify the impact of sharing on throughput and mobility performance. 

\simpletitle{Research Gap:}While the related works discussed above advance our understanding of RAN sharing, from a lab's theoretical setting'', alternative service models adopted by operators to reduce costs, and Türk and Zeydan's analysis of RAN sharing in 4G, they remain limited in several key aspects. In particular, they do not fully address: (i)~Whether the ``theoretical'' economic and performance gains envisioned for RAN sharing translate to real-world commercial deployments, (ii)~How effectively RAN sharing operates in the 5G era, and (iii)~How RAN sharing impacts the user experience in 5G. Our work directly addresses these gaps and contributes to the understanding of commercial 5G networks under spectrum and RAN sharing. 

\section{Concluding Remarks}
\label{s:conclude}

In this paper, we have presented, to the best of our knowledge, the first measurement-based, quantitative analysis of 5G network sharing ``in the wild''. Through a comprehensive measurement campaign in Spain and detailed analysis, we study how commercial spectrum and \ac{ran} sharing affect users in the 5G era. Specifically, we quantitatively study the performance implications of Orange and Vodafone in Spain under \ac{ran} sharing and evaluate how resource management via pooling translates into \ac{qoe} improvements. Our analysis reveals that participating operators benefit equally in terms of throughput from \ac{ran} sharing, although these throughput gains do not always translate into better \ac{qoe} for users. For an operator (Orange in this study), the performance its subscribed users experience increases significantly when \ac{ran} sharing (\ie when utilizing Vodafone's spectrum and \ac{ran} infrastructure) when compared to non-RAN sharing. Interesting, Vodafone users are not penalized in terms of performance when \ac{ran} sharing with Orange users. Although our study focuses on \ac{ran} sharing in Spain, we believe that our insights are applicable to other countries and scenarios, especially in the development of 6G networks and beyond, where (dynamic) spectrum and RAN sharing are key features under consideration.

\bibliographystyle{IEEEtran}
\bibliography{reference}

\end{document}